\documentclass[a4paper,american,pre,preprint,showpacs,showkeys]{revtex4}
\usepackage[T1]{fontenc}
\usepackage[latin9]{inputenc}
\usepackage{amsmath}
\usepackage{graphicx}
\usepackage{amssymb}

\makeatletter

\providecommand{\tabularnewline}{\\}

\usepackage{babel}
\makeatother

\begin{document}

\title{Density of States for a Short Overlapping-Bead Polymer: Clues
to a Mechanism for Helix Formation?}

\author{James E. Magee}

\email{j.magee@manchester.ac.uk}

\affiliation{School of Chemical Engineering and Analytical Science, The University
of Manchester, PO Box 88, Sackville Street, Manchester M60 1QD, United
Kingdom}

\author{Leo Lue}

\affiliation{School of Chemical Engineering and Analytical Science, The University
of Manchester, PO Box 88, Sackville Street, Manchester M60 1QD, United
Kingdom}

\author{Robin A. Curtis}

\affiliation{School of Chemical Engineering and Analytical Science, The University
of Manchester, PO Box 88, Sackville Street, Manchester M60 1QD, United
Kingdom}


\date{\today{}}

\begin{abstract}
The densities of states are evaluated for very short chain molecules
made up of overlapping monomers, using a model which has previously
been shown to produce helical structure. The results of numerical
calculations are presented for tetramers and pentamers. We show that
these models demonstrate behaviors relevant to the behaviors seen
in longer, helix forming chains, particularly, {}``magic numbers''
of the overlap parameter where the derivatives of the densities of
states change discontinuously, and a region of bimodal energy probability
distributions, reminiscent of a first order phase transition in a
bulk system.
\end{abstract}
\maketitle

\section{\label{sec:Introduction}Introduction}

Helices are a common structural motif in biological molecules, from
the $\alpha$-helix in proteins to the culturally iconic double helix
observed in double-stranded DNA. In a living cell, the adoption of
stable helical structures allows these molecules to place functional
groups in specific positions and orientations, and holds the polymer
backbone away from the solvent, protecting it from chemical attack.
The consensus view of helix formation follows the work of Pauling
et al. \citep{pauling}; biological helices are stabilized by orientationally-dependent
hydrogen bonding, with their chirality arising from the chirality
of the polymer molecule. 

Those same properties which make helical molecules so useful in living
cells also make them useful in the context of nanotechnology. Unfortunately,
while our understanding of biological helices is good at explaining
why polypeptides and polynucleotides do form helices, it does not
provide useful prescriptions for developing alternative helix-forming
molecular architectures. To gain the understanding necessary to develop
such prescriptions, many workers have considered {}``reduced models''
\citep{entropichelix,Magee,Maritanetal2000,helixgroundstates,KempChen1998,Varshneyetal2004}
for helix formation, which attempt to capture the underlying physics
of the phenomenon in as simple a manner as possible.

Over recent years, simulation studies of such reduced models have
yielded surprising results. In particular, several polymer models
have been proposed which produce helical structure while interacting
via isotropic potentials \citep{entropichelix,Magee,Maritanetal2000,helixgroundstates};
that is, helix formation without {}``designed-in'' preferred interactions,
with spontaneous chiral symmetry breaking. Maritan et al. \citep{Maritanetal2000,helixgroundstates}
have shown that helices are {}``maximally compact'' structures for
string-like objects. This suggests that helix formation arises from
geometric symmetry breaking, akin to crystallization. A better understanding
of how this symmetry breaking can arise should lead to the better
prescriptions for helix-forming architectures.

In the study of $\alpha$ helix formation in polypeptides, the starting
point is the observation that helices are quasi one dimensional objects,
which can be looked at as a spin chain. The standard approach \citep{doig,ZimmBragg,LifsonRoig,GibbsDiMarzio}
is to attribute amino acid residue conformations to spins, either
H ({}``helix'', that is, capable of forming a hydrogen bond compatible
with a helical structure) or C ({}``coil'' , otherwise). A spin
chain representation is then made up of these states; in the simplest
form \citep{ZimmBragg}, residues which are neighbors along the peptide
backbone interact according only to their H/C attribution and amino
acid type. Modern versions of this approach \citep{doig} include
many-body {}``capping interactions'', which are non-pairwise, non-local
interactions between residues; the strength of these interactions,
however, still depends only upon the residue type and H/C attribution.
Such models have achieved considerable success in helical structure
prediction for polypeptides. For more general helix-forming systems,
the proper attribution of a backbone segment to {}``H'' or {}``C''
type is not clear. However, the success of the spin chain approach
to helix formation in polypeptides suggest that a similar approach
may be fruitful. 

For a linear polymer of spherically symmetric monomers, single monomers
are not the equivalent of amino acid residues for helix formation,
as they have no internal degrees of freedom. From symmetry arguments,
the minimum possible such building block must be a tetramer; helices
break chiral symmetry, and a tetramer is the shortest length chain
which may exhibit chirality. Similarly, the behavior of a pentamer
should contain information on how neighboring chiral centers interact,
and so forth for longer chains.

In this paper, we seek complete enumeration of the partition function
for tetramers and pentamers, using a simple polymer model which has
previously been shown to produce helices \citep{Magee}. This enumeration
is performed using a methodology similar to that followed by Taylor
\citep{Taylor} for short tangent square-well chains. The intention
is to identify the building blocks necessary for helix formation in
longer chains, and the origins of the behaviors which allow helix
formation in longer chains. The methodology and results of this enumeration
are intended as a staging post for the construction of generic spin-chain
models of helix formation

The remainder of the paper is structured as follows. In Sec. \ref{sec:Model},
the polymer model which is to be studied is described. In Sec. \ref{sec:Methods},
the method by which the partition functions for the model are calculated
is described. The results calculated from these partition functions
are described in Sec. \ref{sec:Results}. Finally, in Sec. \ref{sec:Discussion-and-Conclusions},
these results and their implications are discussed.

\section{\label{sec:Model}Model}

The polymer model consists of a linear chain, bond length $l$, of
$N$ hard spherical monomers with diameter $\sigma$. The degree of
overlap between monomers is determined by the reduced parameter $\sigma/l$.
For $\sigma/l=1$, this is the familiar tangent sphere polymer model.
We consider chains with $\sigma/l\geq1$, that is, with overlapping
monomers. Interactions between non-bonded monomers (separation $r$)
are given by an isotropic square-well potential:

\begin{equation}
u(r)=\begin{cases}
\infty & r\leq\sigma\\
-\epsilon & \sigma<r\leq\lambda\sigma\\
0 & \lambda\sigma<r\end{cases}\label{eq:SWpotn}\end{equation}

\noindent where $\lambda$ is the well width (taken as 1.5 in this
work), and the well depth $\epsilon$ sets the energy (and hence temperature)
scale. We follow the protein literature, by denoting interactions
between particles where $\sigma<r<\lambda\sigma$ as \emph{contacts},
and interactions where $r<\sigma$ as \emph{overlaps}. Interactions
between monomers separated by two bonds along the chain are referred
to as 1-3 interactions; interactions for monomers separated by three
bonds are referred to as 1-4 interactions, and so forth.

\begin{figure}
\includegraphics{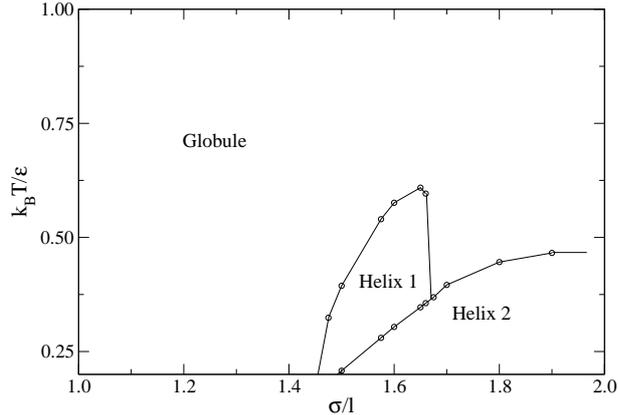}

\caption{\label{fig:phdiag}Schematic phase diagram from simulation for a helix-forming
20mer, as described in the main text. Reproduced from Ref. \citep{Magee}. }

\end{figure}

In previous simulation work, we have used a version of this model
where individual bond lengths were allowed to vary by $\pm10\%$.
It has been suggested that such bond length variation can enhance
the ergodicity of a simulation compared to rigid bonds \citep{homofolding2};
further, this allows the configurational and momentum parts of the
partition function to be factorized. With such bond length fluctuation,
the system has been shown to form helices for 20mers (polymers of
length $N=20$). The observed phase diagram is shown in Fig.~\ref{fig:phdiag};
the system is observed to form two distinct helical phase, {}``helix
1'' (stable at higher temperatures, and with a smaller radius) and
{}``helix 2'' (stable at lower temperatures, and with a larger radius).%
 The following work does \emph{not} include such bond flexibility,
as the extra degree of freedom per bond would make the problem very
much less tractable.

\subsection{Physical Relevance}

With any such {}``reduced model'', however interesting the behaviors,
the question of physical relevance must be answered. The idea of overlapping
monomers is consistent with the Van der Waals radii of atoms in {}``realistic''
potentials such as CHARMM \citep{CHARMM}, where atomic radii are
often larger than the bond length to neighboring atoms. On a larger
scale of approximation, if amino acid residues are approximated by
interacting spheres, the radii of gyration for amino acids can be
larger than their center of mass spacing along the peptide chain.

\begin{figure}
\includegraphics{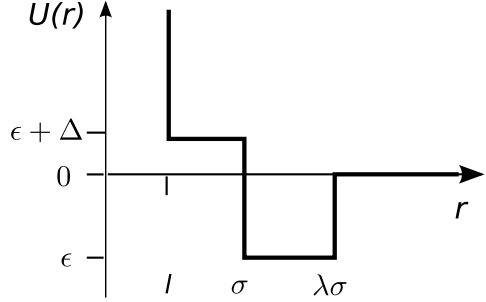}

\caption{\label{fig:coresoftened}A core-softened potential. If the shoulder
height $\Delta\gg kT$, the effective core diameter will be $\sigma$
rather than $l$; hence a chain of monomers interacting via such a
potential with bond length $l$ would act as the overlapping square
well monomer model presented here.}

\end{figure}
Since protein molecules form intra-chain hydrogen bonds, an
interesting parallel can be made to {}``core-softened potentials''
(see Fig.~\ref{fig:coresoftened}), which have been used to study the
anomalous behavior of water
\citep{coresoftened,jaglawater,Stanley}. These isotropic potentials
have a shoulder (diameter $\sigma$) around a repulsive core (diameter
$l$), representing close packed but non-hydrogen bonded pairs, and an
outer well (diameter $\lambda\sigma$) which represents hydrogen
bonding interactions. In a chain of such monomers with bond length
$l$, if the difference between the potential energy of the shoulder
and the potential energy in the minimum is sufficiently larger than
$k_{B}T$, the \emph{effective} repulsive core diameter will be the
shoulder diameter; at low temperatures, a chain of such monomers would
behave as the overlapping square well monomer model presented here,
forming helical structure.

\section{\label{sec:Methods}Methods}

\begin{figure}
\includegraphics{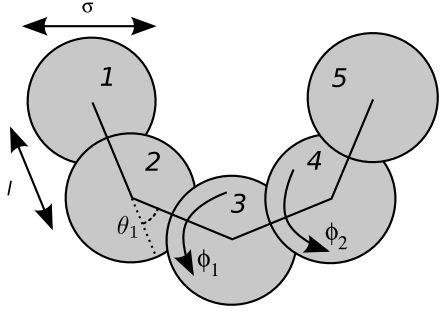}

\caption{\label{fig:Molecule}A cartoon of the model. Five monomers are shown,
of diameter $\sigma$, bond length $l$, with bond angle $\theta_{1}$
and the two dihedral angles $\phi_{1}$ and $\phi_{2}$ indicated.}

\end{figure}

We consider 4- and 5-length polymers of the type described above,
as shown in Fig.~\ref{fig:Molecule}. %
The position of monomer $i$ is denoted by $\mathbf{R}_{i}$. Bond
vectors are defined as $\mathbf{r}_{i}=\mathbf{R}_{i+1}-\mathbf{R}_{i}$.
Separation between monomers $i$ and $j$ is denoted $r_{ij}$. The
bond angle around monomer $i$ is defined as the angle between bond
$\mathbf{r}_{i-1}$ and $\mathbf{r}_{i}$, that is $\cos\theta_{i}=\mathbf{r}_{i-1}\cdot\mathbf{r}_{i}$.
The dihedral (torsional) angle $\phi_{i}$ is defined as the angle
between the planes formed by the vector pairs $\left(\mathbf{r}_{i-1},\mathbf{r}_{i}\right)$
and $\left(\mathbf{r}_{i},\mathbf{r}_{i+1}\right)$, relative to the
\emph{cis} conformation (i.e. $\phi=0$ is \emph{cis}, $\phi=\pi$
is \emph{trans}). We take positive $\phi$ as a right-handed rotation.
This is, however, arbitrary, as the underlying model is achiral. We
do not consider translations and rotations of the entire molecule;
as such, we fix the position of the first monomer, as well as the
plane made by the vectors $\mathbf{r}_{1}$ and $\mathbf{r}_{2}$.
The configurational integral of such an \emph{n}-mer is given by $\mathcal{Z}_{n}$,
defined as:

\begin{eqnarray}
\mathcal{Z}_{n} & = & \left(\prod_{i=1}^{n-3}\int_{-\pi}^{\pi}d\phi_{i}\right)\left(\prod_{j=1}^{n-2}\int_{0}^{\pi}l^{2}\sin\theta_{i}d\theta_{i}\right)\times\nonumber \\
 &  & \quad\exp\left(-\beta E\left(\left\{ \mathbf{R}_{i}\right\} \right)\right)\label{eq:genZ}\end{eqnarray}

\noindent where $E\left(\left\{ \mathbf{R}_{i}\right\} \right)$ is
the total configurational energy for the system (the sum of the pairwise
interactions as Eq. \ref{eq:SWpotn}) and $\beta=1/k_{B}T$, the inverse
temperature.

\subsection{Tetramer\label{sub:Tetramer}}

We initially consider a tetramer. The configurational integral is
given by:

\begin{eqnarray}
\mathcal{Z}_{4} & = & l^{4}\int_{-\pi}^{\pi}d\phi_{1}\int_{0}^{\pi}\sin\theta_{1}d\theta_{1}\int_{0}^{\pi}\sin\theta_{2}d\theta_{2}\times\nonumber \\
 &  & \quad\exp\left(-\beta\left(u\left(r_{13}\right)+u\left(r_{24}\right)+u\left(r_{14}\right)\right)\right)\label{eq:Z4explicit}\end{eqnarray}

Separations are given by:

\begin{equation}
\begin{array}{ccc}
r_{i-1,i+1}^{2} & = & 2l^{2}\left(1+\cos\theta_{i}\right)\\
 & = & l^{2}x_{i}\end{array}\label{eq:xi}\end{equation}

and

\begin{equation}
\begin{array}{ccl}
r_{i-1,i+2}^{2}\left(\theta_{i},\theta_{i+1},\phi_{i}\right) & = & l^{2}\left(\left(1+\cos\theta_{i}+\cos\theta_{i+1}\right)^{2}\right.\\
 &  & \qquad+\sin^{2}\theta_{i}+\sin^{2}\theta_{i+1}\\
 &  & \qquad\left.-2\sin\theta_{i}\sin\theta_{i+1}\cos\phi_{i}\vphantom{\left(\theta\right)^{2}}\right)\\
 & = & l^{2}\left(\left(x_{i}+x_{i+1}-2\right)^{2}/4\right.\\
 &  & \qquad+x_{i}\left(4-x_{i}\right)/4\\
 &  & \qquad+x_{i+1}\left(4-x_{i+1}\right)/4\\
 &  & \qquad-\cos\phi_{i}\sqrt{x_{i}\left(4-x_{i}\right)}\times\\
 &  & \qquad\left.\sqrt{x_{i+1}\left(4-x_{i+1}\right)}/2\right)\\
 & = & l^{2}y_{i}\left(x_{i},x_{i+1},\phi_{i}\right)\end{array}\label{eq:yi}\end{equation}

Since we are working with variables of squared separation, for notational
convenience we also define $a=\left(\sigma/l\right)^{2}$. Physical
bounds for $x_{i}$ are $a\leq x_{i}\leq4$, since a separation of
less than $\sigma$ represents an overlap. It is natural to switch
variables in the configurational integral to separations $x_{i}$,
giving:

\begin{eqnarray}
\mathcal{Z}_{4} & = & (1/4)\int_{-\pi}^{\pi}d\phi_{1}\int_{0}^{4}dx_{1}\int_{0}^{4}dx_{2}\times\nonumber \\
 &  & \quad\exp\left(\vphantom{\left(\sqrt{\phi}\right)}-\beta\left(u\left(l\sqrt{x_{1}}\right)+u\left(l\sqrt{x_{2}}\right)\right.\right.\nonumber \\
 &  & \qquad\left.\left.+u\left(l\sqrt{y_{1}\left(x_{1},x_{2},\phi_{1}\right)}\right)\right)\right)\label{eq:Z4xi}\end{eqnarray}

Since we are working in a square-well system with discretized energies,
it is now convenient to switch to a density-of-states representation:

\begin{equation}
\mathcal{Z}_{4}=\left(1/4\right)\sum_{k=0}^{3}\omega_{4}(k)\exp(\beta\epsilon k)\label{eq:Z4omega}\end{equation}

\noindent where $\omega_{n}(k)$ is the density of states for the
\emph{n}-mer with $k$ contacts. For the tetramer, we can write the
appropriate integrals:

\begin{eqnarray}
\omega_{4}(0) & = & \int d\phi_{1}\int_{\min(4,\lambda^{2}a)}^{4}dx_{1}\int_{\min(4,\lambda^{2}a)}^{4}dx_{2}\times\nonumber \\
 &  & \quad\Theta\left(y_{1}\left(x_{1},x_{2},\phi_{1}\right)-\lambda^{2}a\right)\label{eq:omega40}\end{eqnarray}

\begin{eqnarray}
\omega_{4}(1) & = & 2\int_{-\pi}^{\pi}d\phi_{1}\int_{\min(4,\lambda^{2}a)}^{4}dx_{1}\int_{a}^{\min(4,\lambda^{2}a)}dx_{2}\times\nonumber \\
 &  & \quad\Theta\left(y_{1}\left(x_{1},x_{2},\phi_{1}\right)-\lambda^{2}a\right)\nonumber \\
 &  & +\int_{-\pi}^{\pi}d\phi_{1}\int_{\min(4,\lambda^{2}a)}^{4}dx_{1}\int_{\min(4,\lambda^{2}a)}^{4}dx_{2}\times\nonumber \\
 &  & \quad\left[\Theta\left(y_{1}\left(x_{1},x_{2},\phi_{1}\right)-a\right)\right.\nonumber \\
 &  & \qquad\left.-\Theta\left(y_{1}\left(x_{1},x_{2},\phi_{1}\right)-\lambda^{2}a\right)\right]\label{eq:omega41}\end{eqnarray}

\begin{eqnarray}
\omega_{4}(2) & = & 2\int_{-\pi}^{\pi}d\phi_{1}\int_{\min(4,\lambda^{2}a)}^{4}dx_{1}\int_{a}^{\min(4,\lambda^{2}a)}dx_{2}\times\nonumber \\
 &  & \quad\left[\Theta\left(y_{1}\left(x_{1},x_{2},\phi_{1}\right)-a\right)\right.\nonumber \\
 &  & \qquad-\left.\Theta\left(y_{1}\left(x_{1},x_{2},\phi_{1}\right)-\lambda^{2}a\right)\right]\nonumber \\
 &  & +\int_{-\pi}^{\pi}d\phi_{1}\int_{a}^{\min(4,\lambda^{2}a)}dx_{1}\int_{a}^{\min(4,\lambda^{2}a)}dx_{2}\times\nonumber \\
 &  & \quad\Theta\left(y_{1}\left(x_{1},x_{2},\phi_{1}\right)-\lambda^{2}a\right)\label{eq:omega42}\end{eqnarray}

\begin{eqnarray}
\omega_{4}(3) & = & \int_{-\pi}^{\pi}d\phi_{1}\int_{a}^{\min(4,\lambda^{2}a)}dx_{1}\int_{a}^{\min(4,\lambda^{2}a)}dx_{2}\times\nonumber \\
 &  & \quad\left[\Theta\left(y_{1}\left(x_{1},x_{2},\phi_{1}\right)-a\right)\right.\nonumber \\
 &  & \qquad\left.-\Theta\left(y\left(x_{1},x_{2},\phi_{1}\right)-\lambda^{2}a\right)\right]\label{eq:omega43}\end{eqnarray}

\noindent where $\Theta\left(x\right)$ is the Heaviside step function.
The min terms exist to prevent unphysical limits of integration when
$\lambda^{2}a\geq4$ (in which case 1-3 contacts are {}``always on'').
We note that $\omega_{4}(k)$ is a sum of integrals of the general
form:

\begin{equation}
f_{4}=2\int_{0}^{\pi}d\phi_{1}\int_{x_{1l}}^{x_{1h}}dx_{1}\int_{x_{2l}}^{x_{2h}}dx_{2}\Theta\left(y_{1}\left(x_{1},x_{2}\phi_{1}\right)-h\right)\label{eq:f4}\end{equation}

\noindent where we have used the symmetry of the system to simplify
the $\phi_{1}$ integral. The integrand is non-zero for that region
of $\left(x_{1},x_{2},\phi_{1}\right)$ space for which $y_{1}>h$.
We can solve Eq. \eqref{eq:yi} to find the bound of this space with
respect to $x_{2}$ (or, by symmetry, $x_{1}$) for given $(\phi,h)$,
which we call $x_{c}$:

\begin{eqnarray}
x_{c}\left(x,\phi,h\right) & = & 2\left(\vphantom{\sqrt{\left(h\right)^{2}}}x\left(t\left(4-x\right)+\left(h-1\right)\right)\right.\nonumber \\
 &  & \quad+\mathrm{sgn(\cos(\phi_{1}))}\sqrt{t\left(4-x\right)x}\times\nonumber \\
 &  & \quad\left.\sqrt{\left(2x\left(2t+h-1\right)-tx^{2}-\left(h-1\right)^{2}\right)}\right)\nonumber \\
 &  & /\left(x\left(x+t\left(4-x\right)\right)\right)\label{eq:xcut}\end{eqnarray}

\noindent where we use $t=\cos^{2}\phi$, and $\mathrm{sgn}(x)$ returns
the sign of $x$. Similarly, we also solve for the value of $\phi$
at which $y_{i}=h$ for given $x_{i}$ and $x_{i+1}$, which we denote
$\phi_{c}$:

\begin{equation}
\phi_{c}\left(x_{i},x_{i+1},h\right)=\begin{cases}
\pi & \Phi<-1\\
\arccos\left({\displaystyle \Phi}\right) & \left|\Phi\right|<1\\
0 & \Phi>1\end{cases}\label{eq:phicut}\end{equation}

\noindent where the ratio $\Phi$ is defined as:

\noindent \begin{equation}
\Phi={\displaystyle \frac{2+x_{i}x_{i+1}-2h}{\sqrt{x_{i}\left(4-x_{i}\right)x_{i+1}\left(4-x_{i+1}\right)}}}\label{eq:auxPhi}\end{equation}

We consider the shape of this boundary:

\begin{eqnarray}
\left.\frac{\partial\phi_{c}}{\partial x_{i}}\right|_{x_{i+1}} & = & \frac{2\sqrt{x_{i+1}\left(4-x_{i+1}\right)}}{\sin\phi_{c}\sqrt{x_{i}\left(4-x_{i}\right)}}\times\nonumber \\
 &  & \quad\left(2(1-h)-x_{i}\left(x_{i+1}+(1-h)\right)\right)\label{eq:dphidx}\end{eqnarray}

The only part of this equation which can be negative is the final
bracket. Within the range $a\leq x_{i}\leq4$, $a\leq h\leq\lambda^{2}a$,
$1\leq a<4$ and $\lambda^{2}>1$, it can be shown that $\left.\frac{\partial\phi_{c}}{\partial x_{i}}\right|_{x_{i+1}}$
is non-positive. By symmetry, $\left.\frac{\partial x_{c}}{\partial x}\right|_{\phi}$
is therefore also non-positive. As such, we can write Eq. \eqref{eq:f4}
as:

\begin{equation}
f_{4}=2\int_{0}^{\pi}d\phi_{1}\int_{x_{1l}}^{x_{1h}}dx_{1}\int_{\min\left(x_{2h},\max\left(x_{2l},x_{c}\left(x_{1},\phi,h\right)\right)\right)}^{x_{2h}}dx_{2}\label{eq:f4step1}\end{equation}

The max term picks the larger of the original lower limit ($x_{2l}$)
and the value of $x_{2}$ below which the Heaviside function integrand
in Eq. \eqref{eq:f4} becomes zero. The min prevents the unphysical
result of the lower limit becoming larger than than the upper limit.

We note that, by symmetry, if $x_{1}=x_{c}\left(x_{2},\phi,h\right)$,
then $x_{2}=x_{c}\left(x_{1},\phi,h\right)$. Hence, we can immediately
see that the solution to $x_{2h}=x_{c}\left(x_{1},\phi_{1},h\right)$
with respect to $x_{1}$ is $x_{1}=x_{c}\left(x_{2h},\phi,h\right)$.
For $x_{1}<x_{c}\left(x_{2h},\phi,h\right)$, the upper and lower
limits on the innermost integral are equal, and hence the contribution
to the integral is zero, hence:

\begin{eqnarray}
f_{4} & = & 2\int_{0}^{\pi}d\phi_{1}\int_{\min\left(x_{1h},\max\left(x_{1l},x_{c}\left(x_{2h},\phi,h\right)\right)\right)}^{x_{1h}}dx_{1}\times\nonumber \\
 &  & \quad\int_{\max\left(x_{2l},x_{c}\left(x_{1},\phi,h\right)\right)}^{x_{2h}}dx_{2}\label{eq:f4step2}\end{eqnarray}

Since $x_{c}\left(x,\phi,h\right)$ is a monotonically decreasing
function of $x$ across the range of interest, we can now propagate
the max term in the middle integral out to the dihedral integral:

\begin{eqnarray}
f_{4} & = & 2\int_{\phi_{c}\left(x_{1l},x_{2h},h\right)}^{\pi}d\phi_{1}\int_{x_{1l}}^{x_{1h}}dx_{1}\times\nonumber \\
 &  & \quad\int_{\max\left(x_{2l},x_{c}\left(x_{1},\phi,h\right)\right)}^{x_{2h}}dx_{2}\nonumber \\
 &  & +2\int_{0}^{\phi_{c}\left(x_{1l},x_{2h},h\right)}d\phi_{1}\int_{\min\left(x_{1h},x_{c}\left(x_{2h},\phi,h\right)\right)}^{x_{1h}}dx_{1}\times\nonumber \\
 &  & \quad\int_{\max\left(x_{2l},x_{c}\left(x_{1},\phi,h\right)\right)}^{x_{2h}}dx_{2}\label{eq:f4step3}\end{eqnarray}

\noindent The remaining max and min terms can then be propagated out
in a similar fashion, remembering that $\phi_{c}\left(x,x',h\right)$
is a monotonically decreasing function of $x$ and $x'$ across the
range of interest:

\begin{equation}
\begin{array}{l}
f_{4}=2\int_{\phi_{c}\left(x_{1l},x_{2l}\right)}^{\pi}d\phi_{1}\int_{x_{1l}}^{x_{1h}}dx_{1}\int_{x_{2l}}^{x_{2h}}dx_{2}\\
+2\int_{\max\left(\phi_{c}\left(x_{1l},x_{2h}\right),\phi_{c}\left(x_{1h},x_{2l}\right)\right)}^{\phi_{c}\left(x_{1l},x_{2l}\right)}d\phi_{1}\int_{x_{c}\left(x_{2l}\right)}^{x_{1h}}dx_{1}\int_{x_{2l}}^{x_{2h}}dx_{2}\\
+2\int_{\max\left(\phi_{c}\left(x_{1l},x_{2h}\right),\phi_{c}\left(x_{1h},x_{2l}\right)\right)}^{\phi_{c}\left(x_{1l},x_{2l}\right)}d\phi_{1}\int_{x_{1l}}^{x_{c}\left(x_{2l}\right)}dx_{1}\int_{x_{c}}^{x_{2h}}dx_{2}\\
+2\int_{\phi_{c}\left(x_{1l},x_{2h}\right)}^{\max\left(\phi_{c}\left(x_{1l},x_{2h}\right),\phi_{c}\left(x_{1h},x_{2l}\right)\right)}d\phi_{1}\int_{x_{1l}}^{x_{1h}}dx_{1}\int_{x_{c}}^{x_{2h}}dx_{2}\\
+2\int_{\min\left(\phi_{c}\left(x_{1l},x_{2h}\right),\phi_{c}\left(x_{1h},x_{2l}\right)\right)}^{\phi_{c}\left(x_{1l},x_{2h}\right)}d\phi_{1}\int_{x_{c}\left(x_{2l}\right)}^{x_{1h}}dx_{1}\int_{x_{2l}}^{x_{2h}}dx_{2}\\
+2\int_{\min\left(\phi_{c}\left(x_{1l},x_{2h}\right),\phi_{c}\left(x_{1h},x_{2l}\right)\right)}^{\phi_{c}\left(x_{1l},x_{2h}\right)}d\phi_{1}\int_{x_{c}\left(x_{2h}\right)}^{x_{c}\left(x_{2l}\right)}dx_{1}\int_{x_{c}}^{x_{2h}}dx_{2}\\
+2\int_{\phi_{c}\left(x_{1h},x_{2h}\right)}^{\min\left(\phi_{c}\left(x_{1l},x_{2h}\right),\phi_{c}\left(x_{1h},x_{2l}\right)\right)}d\phi_{1}\int_{x_{c}\left(x_{2h}\right)}^{x_{1h}}dx_{1}\int_{x_{c}}^{x_{2h}}dx_{2}\end{array}\label{eq:f4explimits}\end{equation}

For notational convenience, superfluous arguments to the functions
$\phi_{c}$ and $x_{c}$ have been omitted; that is, $h$ and variables
of integration. There are min and max terms in the dihedral integral
since, without knowing more about the original limits $x_{il}$ and
$x_{ih}$, it is not possible to tell whether $\phi_{c}\left(x_{1h},x_{2l},h\right)<\phi_{c}\left(x_{1l},x_{2h},l\right)$. 

\begin{figure}
\includegraphics{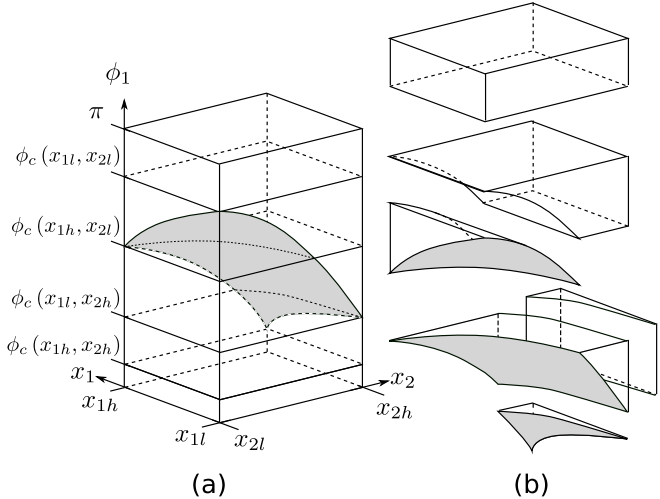}

\caption{Graphical interpretation of the integrals in Eq. \eqref{eq:f4explimits}
when $\phi_{c}\left(x_{1h},x_{2l}\right)>\phi_{c}\left(x_{1l},x_{2h}\right)$.
Part (a) shows the entire space in $x_{1},x_{2},\phi_{1}$; the total
integral is the volume above the gray surface. Part (b) breaks the
volume corresponding to the total integral into the parts listed in
Eq. \eqref{eq:f4explimits}}

\end{figure}

The first integral is trivial. The remaining integrals include a term
$\int x_{c}(x_{1l},\phi_{1},h)d\phi_{1}$; this is analytically tractable,
resulting in terms involving elliptic integrals, but is simpler to
treat numerically. The final two terms include integrals of the form
$\int\int x_{c}(x_{1},\phi_{1},h)dx_{1}d\phi$, which are not analytically
tractable, and are thus treated numerically. 

Calculation of the $f_{4}$ integrals via Eq. \eqref{eq:f4explimits}
allows calculation of the density of states via Eqs. (\ref{eq:omega40}
- \ref{eq:omega43}), from which the equation of state and energy
probability distributions can be determined. Structural information,
in the form of the dihedral angle probability distributions, is also
easily accessible. The dihedral density of states $g_{4}(\phi_{1},k)$
is given by integrals as Eqs. (\ref{eq:omega40} - \ref{eq:omega43})
without the integral over the dihedral angle. For the tetramer, these
give tractable though lengthy analytic forms. The probability of observing
a given dihedral angle is then given by:

\begin{equation}
P\left(\phi_{1};T\right)=\sum_{k=0}^{4}g_{4}\left(\phi_{1},k\right)\exp(\beta\epsilon k)/\mathcal{Z}_{4}\label{eq:P(phi)}\end{equation}

\subsection{Pentamer\label{sub:Pentamer}}

An equivalent procedure may be carried out for a pentamer. We first
introduce the 1-5 separation, $z_{i}\left(x_{i},x_{i+1},x_{i+2},\phi_{i},\phi_{i+1}\right)=r_{i-1,i+3}^{2}/l^{2}$:

\begin{eqnarray}
z_{i} & = & x_{i}+x_{i+2}-\sqrt{x_{i+1}\left(4-x_{i+1}\right)}\times\nonumber \\
 &  & \quad\left(\cos\phi_{i}x_{i+2}\sqrt{x_{i}\left(4-x_{i}\right)}\right.\nonumber \\
 &  & \qquad\left.\cos\phi_{i+1}x_{i}\sqrt{x_{i+2}\left(4-x_{i+2}\right)}\right)/4\nonumber \\
 &  & +\left(x_{i+1}-2\right)\left(\vphantom{\sqrt{\left(x_{i}^{i}\right)}}x_{i}x_{i+2}-\cos\phi_{i}\cos\phi_{i+1}\times\right.\nonumber \\
 &  & \qquad\left.\sqrt{x_{i}\left(4-x_{i}\right)x_{i+2}\left(4-x_{i+2}\right)}\right)/4\nonumber \\
 &  & +\sin\phi_{i}\sin\phi_{i+1}\times\nonumber \\
 &  & \quad\sqrt{x_{i}\left(4-x_{i}\right)x_{i+2}\left(4-x_{i+2}\right)}/2\label{eq:zi}\end{eqnarray}

\noindent where the arguments to $z_{i}$ have been omitted. The partition
function for the pentamer is given by:

\begin{equation}
\mathcal{Z}_{5}=(1/8)\sum_{k=0}^{6}\omega_{5}(k)\exp(\beta\epsilon k)\label{eq:Z5}\end{equation}

Equivalent expressions to Eqs. (\ref{eq:omega40}-\ref{eq:omega43})
are simple to construct, using the equivalent form to Eq. \eqref{eq:f4}:

\begin{eqnarray}
f_{5} & = & 2\int_{0}^{\pi}d\phi_{1}\int_{-\pi}^{\pi}d\phi_{2}\int_{x_{1l}}^{x_{1h}}dx_{1}\int_{x_{2l}}^{x_{2h}}dx_{2}\int_{x_{3l}}^{x_{3h}}dx_{3}\times\nonumber \\
 &  & \quad\Theta\left(y_{1}-h_{1}\right)\Theta\left(y_{2}-h_{2}\right)\Theta\left(z_{1}-h_{3}\right)\label{eq:f5}\end{eqnarray}

\noindent where we have suppressed the arguments of $y_{1}$, $y_{2}$
and $z_{1}$ for notational ease. This integral is constructed (without
loss of generality) such that $\phi_{1}$ is \emph{always} right-handed.
Explicit bounds of integration due to 1-4 interactions can be treated
in the same manner as for the tetramer case. Bounds for the $x_{1}$
and $x_{2}$ integrals as a function of $\phi_{1}$, and for the $x_{2}$
and $x_{3}$ integrals as a function of $\phi_{2}$, are determined
exactly as Eq. \eqref{eq:f4explimits}. This leads to single ranges
of integration for $x_{1}$ and $x_{3}$, and two sets of ranges of
integration for $x_{2}$. The proper range of integration over $x_{2}$
is then the overlap of these two ranges. Explicitly treating the bounds
of integration due to 1-5 interactions is not trivial, and as such
the resulting integral is treated numerically. Dihedral probability
distributions $P\left(\phi_{1},\phi_{2},T\right)$ can be calculated
from dihedral densities of states $g_{5}\left(\phi_{1},\phi_{2},k\right)$
in an analogous manner to the tetramer.

\section{\label{sec:Results}Results}

Using the results presented in Sec. \ref{sec:Methods}, we have evaluated
the full partition functions for tetramers and pentamers. Results
for the tetramer have been calculated with the Mathematica symbolic
algebra package, using Gauss-Kronrod numerical integration. Results
for the pentamer have been calculated using ten-point Gauss-Legendre
quadrature \citep{NRC}. Both methods of integration have been checked
by comparison against the tangent chain results presented by Taylor
\citep{Taylor}. The pentamer results have been verified against short
Monte Carlo simulations (data not shown).

\subsection{Tetramer}

To validate the method, we compare our calculated densities of states
for tetramer tangent square well chains ($\sigma/l=1$) to those presented
by Taylor \citep{Taylor}. These results are shown in Table \ref{tab:4mertangentDoS}.%
\begin{table}
\caption{\label{tab:4mertangentDoS}Comparison of the densities of
  states for a square well tetramer chain, $\sigma/l=1.0$ and
  $\lambda\sigma=1.5$ calculated in this work ($\omega_{4}(k)/4l^{4}$,
  with the factor $1/4$ (as described in section \ref{sec:Methods})
  and by Taylor \citep{Taylor} ($\omega_{4}^{(Taylor)}(k)$, or
  $g_{4}^{(k)}$ in the original terminology).  Suppression of the
  unimportant multiplicative factor of $8\pi$ in the work of Taylor
  leads to the difference in the values; it can be seen that including
  this factor, the values differ only in the fourth and fifth
  significant figure.}

\begin{tabular}{cccc}
\hline\hline
$k$ & $\omega_{4}(k)/4l^{4}$ & $\omega_{4}^{(Taylor)}(k)$ & $\omega_{4}(k)/\left(32\pi\omega_{4}^{(Taylor)}(k)\right)$\tabularnewline
\hline 
0 & 4.78131 & 0.19029 & 0.999750\tabularnewline

1 & 5.59121 & 0.22247 & 0.999987\tabularnewline

2 & 2.42528 & 0.09650 & 0.999986\tabularnewline

3 & 0.62013 & 0.02467 & 1.000170\tabularnewline
\hline\hline
\end{tabular}
\end{table}
It can be seen that the results are equivalent to four significant
figures aside from an unimportant multiplicative factor. The method
of Taylor does not use explicit limits of integration, instead numerically
integrating the Heaviside functions in Eq. \eqref{eq:f4}; strictly,
the method presented here should be more accurate, though these results
suggest the difference is not significant. 

\begin{figure}
\includegraphics{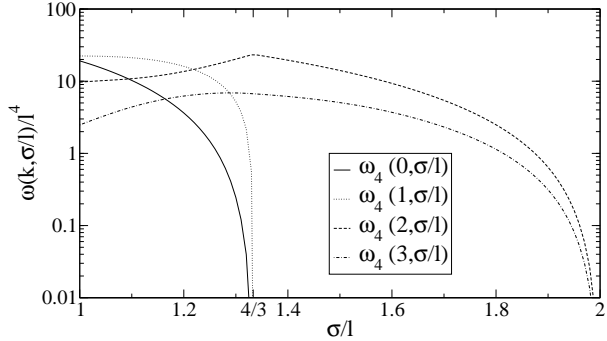}

\caption{\label{fig:4merDoS}Densities of states for tetramers plotted against
$\sigma/l$.}

\end{figure}

The calculated densities of states as a function of $\sigma/l$ are
shown in Fig.~\ref{fig:4merDoS}.%
The densities of states for the $k=0$ and $k=1$ states are zero
for $\sigma/l\geq4/3$. For overlaps greater than this {}``magic
number'', 1-3 interactions become {}``always on'' --- that is,
$r_{i-1,i+1}\leq\lambda\sigma$ for all values of $\theta_{i}$ with
$\lambda=3/2$. This also gives rise to a kink (discontinuity in the
derivative) of $\omega_{4}(2,\sigma/l$). This is because at $\sigma/l=4/3$,
the first term in Eq. \eqref{eq:omega42} (which refers to the density
of states for tetramers with a single 1-3 contact and a 1-4 contact)
becomes zero, as the limits on the $x_{1}$ integral become equal. 

As $\sigma/l\rightarrow2$, the polymer becomes increasingly rigid,
and the available conformational space vanishes. The calculated densities
of states show the correct behavior at this limit.

\begin{figure}
\includegraphics{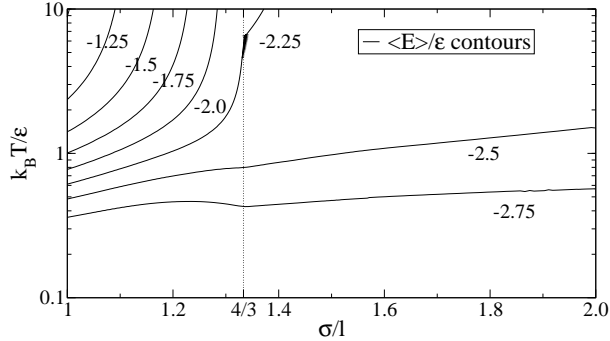}
\caption{\label{fig:4merE}Ensemble average energies $\left\langle
  E\right\rangle /\epsilon$ for tetramers plotted against $\sigma/l$
  and temperature $T$. Solid lines show energy contours at the labeled
  value. Note the discontinuities in the slope of the energy contours
  at $\sigma/l=4/3$ (shown by the dotted line).}
\end{figure}

\begin{figure}
\includegraphics{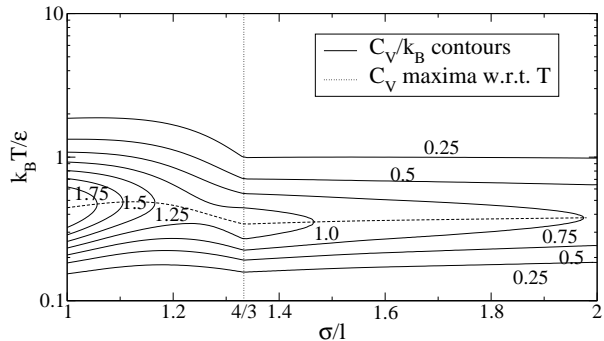}
\caption{\label{fig:4merCv}Configurational heat capacity $C_{V}$ for tetramers
plotted against $\sigma/l$ and temperature. Solid lines show contours
at the labeled value. The dashed line shows the line of maxima in
$C_{V}$ with respect to temperature $T$. Note the discontinuities
in the slope of the heat capacity contours at $\sigma/l=4/3$ (shown
by the dotted line).}
\end{figure}

Properties calculated from these densities of states are shown in
Figs.~\ref{fig:4merE} (energy) and \ref{fig:4merCv} (heat capacity).

At $\sigma/l=4/3$, the slope of the energy and heat capacity contours
show discontinuities in their derivatives. As such, the derivatives 
$\left(\partial U/\partial\sigma\right)_{T}$ and 
$\left(\partial C_{V}/\partial\sigma\right)_{T}$ have singularities at 
$\sigma/l=4/3$, however, these are not physically meaningful response 
functions. In simulated systems \citep{Magee}, bond lengths are not rigid,
 and bond length fluctuations will have the effect of {}``smoothing out'' 
the discontinuity.

Though the tetramer does not show any other discontinuities, it does
show a line of maxima in heat capacity with respect to temperature.
We follow Taylor \citep{Taylor} and Zhou, et al.\ \citep{homofolding2}
in ascribing these maxima to collapse of the tetramer into compact
conformations. The strength of these maxima can be seen to decrease
with increase in $\sigma/l$. Further, the line of maxima shows
re-entrance with respect to $\sigma/l$, with the temperature at which
heat capacity is maximal itself having a maximum with respect to
$\sigma/l$ at a point below $\sigma/l=4/3$.

\begin{figure*}
\includegraphics[scale=0.5]{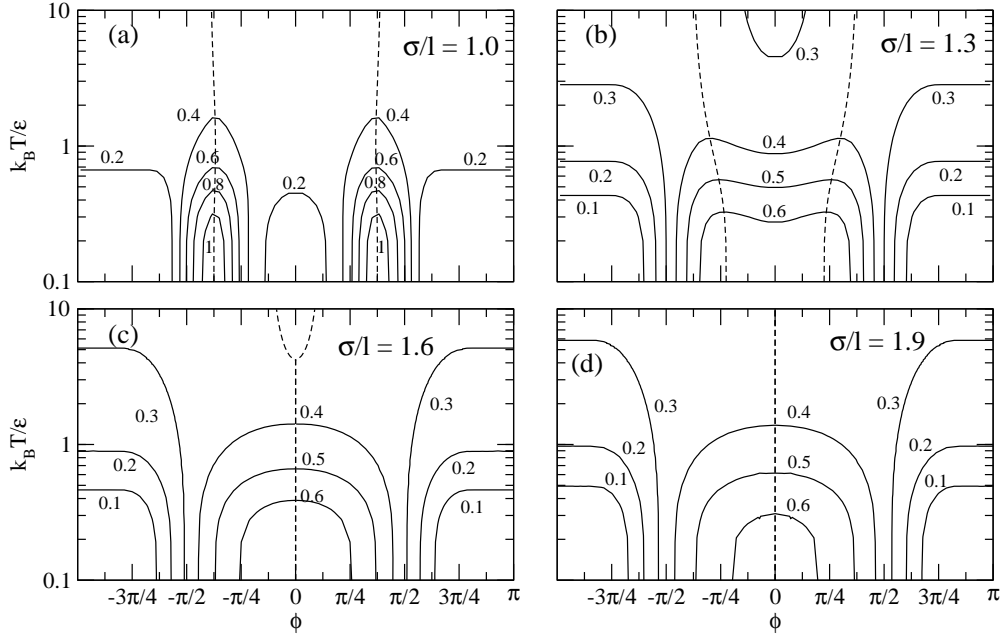}

\caption{\label{fig:4merPphis}Contour plots of $P\left(\phi_{1};T\right)$
for (a) $\sigma/l=1.0$, (b) $\sigma/l=1.3$, (c) $\sigma/l=1.6$
and (d) $\sigma/l=1.9$. Solid lines show contours at the labeled
value. Dashed lines show maxima in $P\left(\phi_{1};T\right)$. }

\end{figure*}

\begin{figure}
\includegraphics{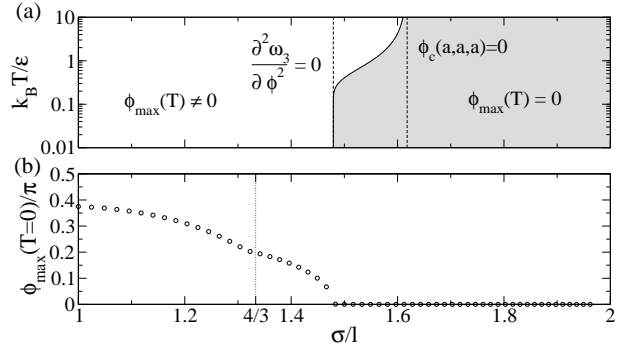}

\caption{\label{fig:phimax} (a) Behavior of $P(\phi=0;T)$ with respect to
$\sigma/l$; to the left of the solid line, $P(\phi=0)$ is a minimum,
whilst to the right, $P(\phi=0)$ is a maximum. Directly on the line,
$P(\phi=0)$ is a point of inflection. Dashed lines indicate the upper
and lower bounds in $\sigma/l$ of the line. (b) The behavior of the
maximum of $P(\phi_{1};T=0)$ (that is, in the ground state) with
respect to $\sigma/l$. There is a kink in the line at $\sigma/l=4/3$,
indicated by the dotted line.}

\end{figure}

Representative results for the torsional behavior of the tetramer
are shown in Fig.~\ref{fig:4merPphis},%
where we show the probability $P(\phi_{1};T)$ for four values of
$\sigma/l$. At low values of the overlap ($\sigma/l\lesssim1.48$),
we see maxima in $P(\phi_{1};T)$ for non-zero $\phi_{1}$ at all
temperatures, with the maxima becoming stronger and moving closer
to zero (\emph{cis} conformation) as temperature decreases. For intermediate
values of overlap ($1.48\lesssim\sigma/l<\sqrt{(3+\sqrt{5})/2}$),
weak maxima in $P(\phi_{1};T$) are seen for non-zero $\phi_{1}$
at high temperatures; however, the most probable conformation becomes
\emph{$\phi_{1}=0$ }(\emph{cis }conformation) at low temperature.
For large values of overlap ($\sigma/l>\sqrt{(3+\sqrt{5})/2}$), $P(\phi_{1};T)$
has only a single maximum at $\phi_{1}=0$ for all temperatures. The
points separating the two regimes ($\max\left(P\left(\phi_{1};T\right)\right)=0$
and $\neq0$) can be calculated analytically, as the points at which
$\left.{\displaystyle \frac{\partial P(\phi_{1};T)}{\partial\phi_{1}}}\right|_{\phi_{1}=0}=0$.
The calculated line in overlap-temperature space is shown in Fig.~\ref{fig:phimax} (a)%
. The upper limit of this line is the value of $\sigma/l$ at which
$\phi_{c}\left(x_{i}=x_{i+1}=h=a\right)=0$. For values of overlap
equal to or larger than this, it is not possible for the polymer to
exhibit 1-4 overlaps, and there is no steric hindrance to $\phi_{1}=0$
states, which are the points of closest 1-4 approach. The lower limit
of this region occurs at the point where the maximum of $P(\phi_{1};T=0)$
(see Fig.~\ref{fig:phimax} (b)) becomes zero. The value of $\sigma/l$
at this limit does not admit a simple interpretation or expression.

\subsection{Pentamer}

\begin{figure}
\includegraphics{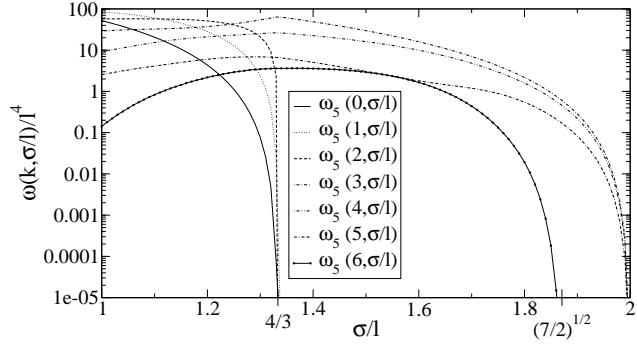}

\caption{\label{fig:5merDoS}Densities of states for pentamers plotted against
$\sigma/l$. }

\end{figure}

A comparison between the densities of states calculated here for pentamer
tangent square well chains with those presented by Taylor is provided
in Table \ref{tab:5mertangentDoS}.%
\begin{table}
\caption{\label{tab:5mertangentDoS}Comparison of the densities of states for
a square well tetramer chain, $\sigma/l=1.0$ and $\lambda\sigma=1.5$
calculated in this work ($\omega_{5}(Ek)/4l^{4}$, with the factor
$1/4$ as described in section \ref{sec:Methods}) and by Taylor ($\omega_{5}^{(Taylor)}(k)$,
or $g_{5}^{(k)}$ in the original terminology) . Suppression of an
unimportant multiplicative factor of $16\pi^{2}$ in the work of Taylor
leads to the difference in the values; it can be seen that including
this factor, the differences in the values are negligible.}
\begin{tabular}{cccc}
\hline\hline  
k & $\omega_{5}(k)/4l^{4}$ & $\omega_{4}^{(Taylor)}(k)$ & $\omega_{4}(k)/\left(64\pi^{2}\omega_{4}^{(Taylor)}(k)\right)$\tabularnewline
\hline
0 & 12.963393 & 0.08206 & 1.000386\tabularnewline

1 & 21.368822 & 0.13531 & 1.000071\tabularnewline

2 & 14.300928 & 0.09057 & 0.999908\tabularnewline

3 & 7.300283 & 0.04626 & 0.999342\tabularnewline

4 & 2.284838 & 0.01447 & 0.999924\tabularnewline

5 & 0.0633290 & 0.004012 & 0.999590\tabularnewline

6 & 0.035383 & 0.0002222 & 1.008395\tabularnewline
\hline\hline 
\end{tabular}
\end{table}
Results are equivalent to four significant figures. The calculated
densities of states as a function of $\sigma/l$ are shown in Fig.~\ref{fig:5merDoS}.%
Once again, we see the highest energy densities of states going to
zero at $\sigma/l=4/3$ as 1-3 interactions become {}``always on'',
combined with a kink in the density of states for the highest remaining
energy. All densities of states tend to zero as $\sigma/l\rightarrow2$,
where the available conformational space becomes zero. There are two
further behaviors, not seen in the tetramer. The most obvious is that
the density of the lowest energy state $\omega_{5}(6)$ becomes zero
at $\sigma/l=\sqrt{7/2}$. For values of overlap larger than this,
the pentamer has become so stiff that it cannot bend back on itself
far enough to make 1-5 contacts.

\begin{figure}
\includegraphics{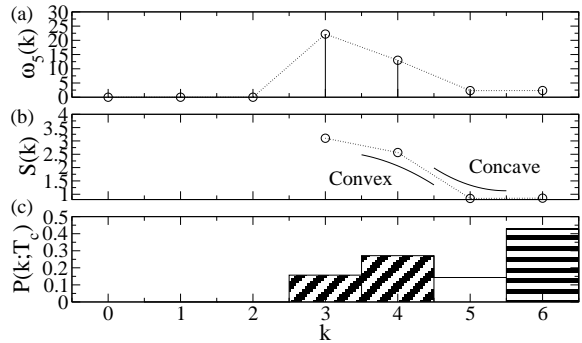}

\caption{\label{fig:convexity}Concavity in the entropy and bimodal energy
probability distribution for $\sigma/l=1.55$. (a) The density of
states $\omega_{5}(k)$. (b) The entropy as a function of $k$; note
the concavity at $k=5$. (c) The probability distribution function
$P(k,T)$ at the {}``state coexistence'' temperature. The function
is bimodal, and the total weights of the two {}``states'' ($k<5$
(diagonal shading) and $k>5$ (horizontal shading)) are equal. Dashed
lines serve as a guide to the eye.}

\end{figure}

\begin{figure}
\includegraphics{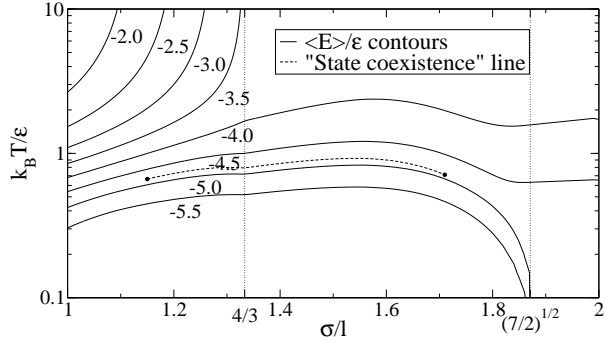}

\caption{\label{fig:5merE}Ensemble average energies $\left\langle E\right\rangle /\epsilon$
for pentamers plotted against $\sigma/l$ and temperature $T$. Solid
lines show energy contours at the labeled value. The dashed line shows
the {}``state coexistence'' line. Note the discontinuities in the
slope of the energy contours at $\sigma/l=4/3$ and $\sigma/l=\sqrt{7/2}$(shown
by dotted lines).}

\end{figure}

\begin{figure}
\includegraphics{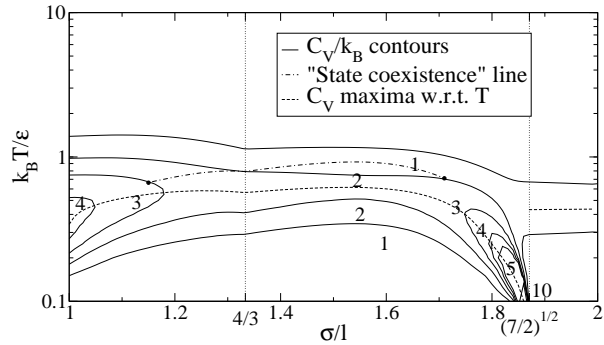}

\caption{\label{fig:5merCv}Configurational heat capacity $C_{V}$ for pentamers
plotted against $\sigma/l$ and temperature. Solid lines show contours
at the labeled value. The dashed line shows the line of maxima in
$C_{V}$ with respect to temperature $T$. The dot-dashed line shows
the {}``state coexistence'' line --- the position of the end points
of this line near a contour line is purely coincidental. Note the
discontinuities in the slope of the heat capacity contours at $\sigma/l=4/3$
and $\sigma/l=\sqrt{7/2}$ (shown by dotted lines).}

\end{figure}

A further interesting behavior is observed at intermediate values
of $\sigma/l$ where the ground state $\omega_{5}(6)$ becomes the
same order of magnitude as $\omega_{5}(5$). Indeed, for $1.53\lesssim\sigma/l\lesssim1.56$,
$\omega_{5}(6)>\omega_{5}(5)$. This gives rise to a concavity in
the entropy $S(k)=k_{B}\ln\omega(k)$ of the system with respect to
energy at $E=-5\epsilon$, which can be studied using the discrete
analog to the second derivative, $S''(k)=\left(S(k+1)-2S(k)+S(k-1)\right)$;
the function is concave if $S''(k)$ is negative. The concavity results
in a bimodal probability distribution function $P(E;T,\sigma/l)$
(illustrated in Fig.~\eqref{fig:convexity})%
. In analogy to the study of phase transitions, we find the line of
temperatures at which the two peaks of these bimodal probability distributions
have equal weight - a line of {}``state coexistence''. This line
is plotted alongside the data in Figs. \ref{fig:5merE} (energy)%
 and \ref{fig:5merCv} (heat capacity),%
 and runs from $\sigma/l\approx1.14$ to $\sigma/l\approx1.72$. These
end points are at non-zero temperature, and occur where the curvature
of the free energy at $E=-5\epsilon$ becomes zero. The end points
are not associated with heat capacity divergences.

The thermodynamic data shows the expected discontinuities in the slope
of energy and heat capacity contour at the {}``magic numbers'' $\sigma/l=4/3$
and $\sigma/l=\sqrt{7/2}$. The lower magic number corresponds to
the loss of high energy states, as for the tetramer. The larger magic
number, corresponding to the loss of the $k=6$ state, gives a discontinuity
in the energy at zero temperature (from $E=-6\epsilon$ to $E=-5\epsilon$).
As for the tetramer, bond length fluctuations in real systems would
act to smooth out these discontinuities in real systems. 

The pentamer also shows a line of heat capacity maxima, which lies
at lower temperature than the {}``state coexistence'' line. Both
these lines show a discontinuity in slope at $\sigma/l=4/3$. Both
lines are doubly reentrant, showing one maximum below $\sigma/l=4/3$,
and another above $\sigma/l=4/3$. The line of heat capacity maxima
connects with the discontinuity in energy at $\sigma/l=\sqrt{7/2}$.

\begin{figure*}
\includegraphics{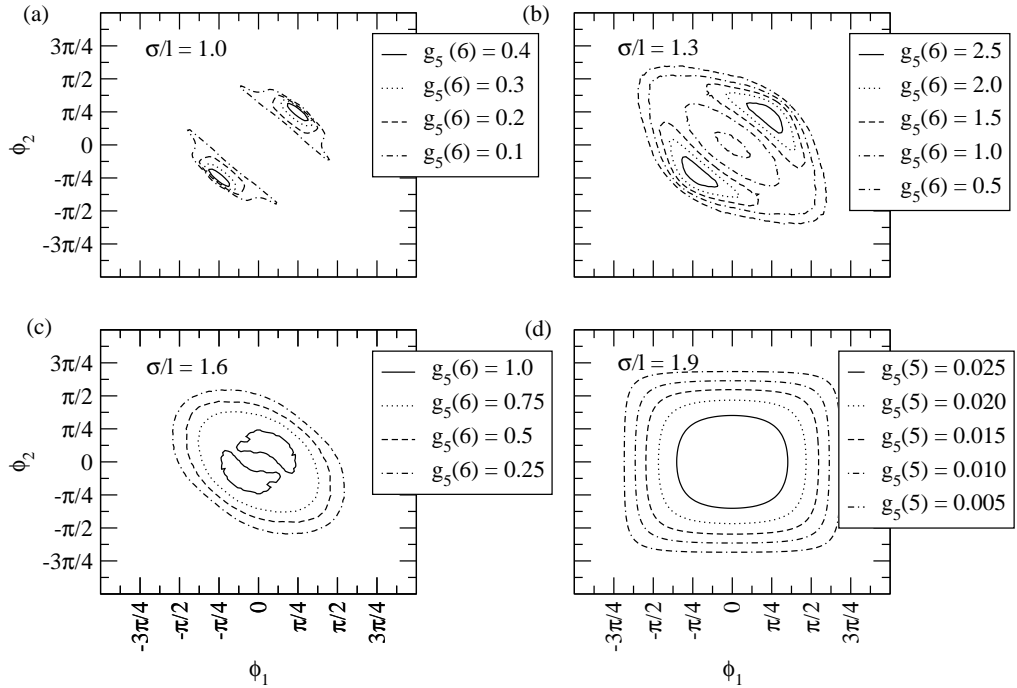}

\caption{\label{fig:gsphis}Ground state dihedral densities of states $g_{5}\left(\phi_{1},\phi_{2},6\right)$
for (a) $\sigma/l=1.0$, (b) $\sigma/l=1.3$, and (c) $\sigma/l=1.6$,
and (d) $g_{5}\left(\phi_{1},\phi_{2},5\right)$ for $\sigma/l=1.9$;
the ground state is $-5\epsilon$ for $\sigma/l=1.9$. Lines show
contours as denoted in figure legends. }

\end{figure*}

The dihedral behavior of the pentamer at zero temperature (ground
state) is shown in Fig.~\ref{fig:gsphis},%
in four representative plots of the ground state dihedral densities
of states $g_{5}\left(\phi_{1},\phi_{2},6\right)$ and $g_{5}\left(\phi_{1},\phi_{2},5\right)$.
These are equivalent to unnormalized dihedral probability distributions
for the system at $T=0$. We see that, for the three lowest values
of $\sigma/l$, the probabilities are peaked at points on the $\phi_{1}=\phi_{2}$
diagonal; the dihedrals prefer to take the same sign. This continues
to the {}``magic number'' $\sigma/l=\sqrt{(3+\sqrt{5})/2}$, where,
as for the tetramer, 1-4 overlaps can no longer occur, and steric
effects no longer prevent \emph{cis} conformations. As the degree
of overlap tends to this number, and the amount of steric interference
decreases, the maxima move closer to $\phi_{1}=\phi_{2}=0$. For $\sigma/l\geq\sqrt{(3+\sqrt{5})/2}$,
the probability distributions become unimodal at $\phi_{1}=\phi_{2}=0$.
This should be compared with the behavior for the tetramer (see Fig.~\ref{fig:phimax}), where $\phi_{max}(T=0)$ is zero for $\sigma/l\gtrsim1.48$;
this effect is due to the additional steric interference from 1-5
overlaps.

\section{\label{sec:Discussion-and-Conclusions}Discussion and Conclusions}

In the previous section, it has been shown that the tetramer and pentamer
show a rich and surprising range of behaviors. Specifically, these
are {}``magic numbers'' of the overlap $\sigma/l$ where the derivatives
of the densities of states change discontinuously, maxima in specific
heat with respect to temperature, and a region of bimodal energy probability
distributions, reminiscent of a first-order transition in bulk systems.
In general, the behavior of long polymer chains cannot be directly
inferred from the behavior of very short chains such as those studied
in this work. If, however, interactions between monomers widely spaced
along a chain can be neglected, the behavior of very short chains
can be used as a basis for a spin chain model. Such interactions may
be neglected when chains become very stiff (at, for e.g., large values
of $\sigma/l$, or after helix formation). In this case, the behavior
of the very short chains may be considered the {}``building block''
for the behavior of longer chains.

The {}``magic numbers'' which are observed correspond to discontinuous
changes in the derivatives of the densities of states. At $\sigma/l=4/3$,
1-3 contacts become {}``always on'' and high energy densities of
states become zero. At $\sigma/l=\sqrt{(3+\sqrt{5})/2}$, the chain
becomes so stiff that it cannot bend back upon itself far enough for
1-4 overlaps to occur. Similarly, at $\sigma/l=\sqrt{7/2}$, the chain
becomes so stiff that 1-5 contacts can no longer occur, and the ground
state for the pentamer is lost. These discontinuities in the densities
of states are associated with discontinuities in the energy and compressibility
with respect to the parameter $\sigma$. The {}``magic numbers''
are similar in principle to the {}``cut-off'' $\lambda$ values
noted by Taylor \citep{Taylor} for tangent chains --- this work has
not examined the effects of changing the well width parameter $\lambda$,
but it is obvious that the values of these {}``magic numbers'' will
depend upon that parameter, and that {}``cut-off'' values of $\lambda$
will also exist for this model. As has been noted above, the discontinuities
across lines of constant $\sigma/l$ in this system will be smoothed
in simulations with variable bond length, however, the effects should
still be visible. We particularly note the sudden loss of stability
of the {}``helix 1'' phase at $\sigma/l\approx1.675$ in previous
simulation work \citep{Magee} (see Fig.~\ref{fig:phdiag}). Given
the 10\% bond length fluctuation allowed in those simulations, this
loss of stability may coincide with the magic number at $\sigma/l=\sqrt{(3+\sqrt{5})/2}\approx1.618$,
suggesting that the more tightly wound {}``helix 1'' phase is stabilized
by steric interference of 1-4 contacts. This supposition is supported
by the observed loss of double-peaked dihedral angle probability distributions
for overlaps above this {}``magic number'', suggesting that the
more loosely wound {}``helix 2'' phase is stabilized by steric interference
between monomers spaced further along the chain.

The low temperature maxima in the specific heat for these short polymers
appear to be a continuation of the specific heat maxima observed for
short tangent chains \citep{homofolding2,Taylor}. We follow these
previous works in interpreting these maxima as signatures of collapse
to close-packed, low energy conformations. This interpretation appears
confirmed by the presence of bimodal energy probability distributions
for the pentamer, with a line of {}``state coexistence'' which roughly
parallels the line of maxima. 

For the tetramer, the line of specific heat maxima shows re-entrance
below $\sigma/l=4/3$, having a maximum with respect to temperature.
For the pentamer, both the line of specific heat maxima and of {}``state
coexistence'' are doubly reentrant, showing maxima below and above
$\sigma/l=4/3$. The re-entrance of the {}``state coexistence''
line can be easily explained by reference to the densities of states
shown in Fig.~\ref{fig:5merDoS}. Consider the system for $\sigma/l\geq4/3$.
For overlaps just above this point, the ground state density of states
(the entropy of the low energy state) is increasing while all other
densities of states are decreasing with increasing overlap. Hence,
the low energy state becomes more stable, and coexistence moves to
higher temperature. The ground state density of states soon begins
to decrease, but as long as it is decreasing \emph{more slowly} than
the higher energy density of states, its stability continues to increase.
However, on closer approach to $\sigma/l=\sqrt{7/2}$, the ground
state density of states decreases \emph{faster} than the higher energy
densities of states, and stability decreases. The same argument holds
for the line when $\sigma/l<4/3$. If we interpret the maximum in
heat capacity as a result of \emph{structural competition} between
the ground state and higher energy states (following Stanley, et al.\
\citep{Water}), we can make the same argument for the re-entrance
in the lines of maxima for both the tetramer and pentamer. Physically,
increasing the overlap of the chain makes configurations with lower
energy (more contacts) more likely at first (as monomers are {}``drawn
into'' each other's square wells), but then begins to cut into these
low energy states as the chain becomes too stiff to bend back upon
itself and make contacts. We attribute the re-entrance of the stability
of the {}``helix 1'' phase in previous work to this same competition
between effects.

Though it seems reasonable to attribute the behavior of the phase
boundary between the {}``helix 1'' and globule phases to effects
seen in the pentamer, it should be noted that the state coexistence
seen in the pentamer is represents collapse of the pentamer, rather
than helix formation. Though the dihedral probability distributions
shown in Fig.~\ref{fig:gsphis} do show double peaks at non-zero dihedral
angles, this is not a sufficient criterion for helicity. The cross-correlation
coefficient of these distributions is not significantly above zero;
the total statistical weight associated with dihedrals away from the
peaks is still large enough to outweigh the correlated peaks. However,
the clear double peaked structure does suggest that the physics necessary
for helix formation \emph{is} contained in these simple, small systems,
particularly in the steric interference due to 1-4 overlaps.

While these results appear to clarify certain behaviors observed in
simulations, they do raise further questions. Under the interpretation
we have offered here, the nature of the {}``helix-2'' phase is unclear;
this phase is observed to be stable up to $\sigma/l=1.9$ in simulation
\citep{Magee}, where the chain is too stiff for 1-5 overlaps to be
the root of the observed chirality. Further, the question of how the
helix transition connects (or does not connect) to the crystallization-like
transition observed in simulations for the tangent chain system remains
unresolved. Follow-up work, developing a spin chain model for helix
formation using the results presented here, is underway; it is hoped
that this approach will shed light upon these questions. 

\begin{acknowledgments}
This work is supported by the EPSRC (grant reference EP/D002753/1).
James Magee would like to thank Dr. Richard Blythe for interesting
discussions.
\end{acknowledgments}
\bibliographystyle{apsrev}

\end{document}